# Understanding Computer Networks: A Comprehensive Overview of Types, Configurations, and the OSI Model

Priyanshu Tyagi

Rahul

Nishant Priyadarshi

Under the guidance of:

Dr. Jayasheela.C.S

Dept of ISE,

BIT.

## ABSTRACT

Computer networks have evolved into an essential component of modern society, facilitating the seamless sharing and dissemination of digital information. This paper explores the fundamental concepts of networking, focusing on the transformative impact of intranets and internets. Intranets, in particular, have emerged as indispensable tools for businesses, enabling them to efficiently manage and distribute information. The rapid adoption of intranets underscores their role in maintaining competitiveness in today's fast-paced business environment. This article provides an overview of networking fundamentals, emphasizing the significance of intranets and internets in modern computing.

## INTRODUCTION

Networking is the backbone of modern computing, enabling communication between programs running on physically distant machines. A computer network is a collection of interconnected computers that exchange data and share resources. This exchange is facilitated by protocols, which define the rules and conventions for communication over various media. Networks can range from simple setups connecting a few computers to complex infrastructures spanning continents. Understanding the types of network configurations, such as peer-to-peer and client/server networks, is essential for designing efficient and scalable network solutions.

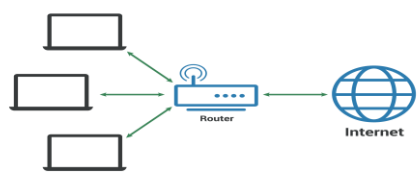

Fig 1: Local Area Network

# LITERATURE REVIEW

## 1. Types of Network Configuration

### i) Peer to Peer Networking

Peer-to-peer networks are typically implemented in environments with fewer than ten computers and where stringent security measures are not critical. In such networks, all computers are considered equal, or "peers," communicating with each other on an equal footing. This means that files can be easily shared across the network, and all computers can access devices like printers or scanners connected to any one computer. In a peer-to-peer network, each computer functions both as a client and a server, allowing for decentralized sharing of resources. This architecture is depicted in Figure 2, illustrating how computers are interconnected in a peer-to-peer network.

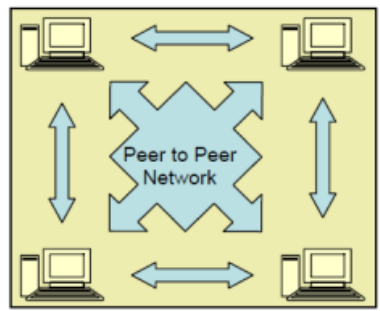

Fig 2: Peer to Peer Networking

### ii) Client/Server Networks

Client/server networks are better suited for larger networks where centralized control and management are important. In a client/server network, a central computer, known as the "server," serves as the storage location for files and applications shared across the network. Typically, the server is a high-performance computer capable of handling multiple client requests simultaneously. The server also manages network access, controlling which computers, known as "clients," can access specific resources on the network. Access to the server is usually restricted to the network administrator, ensuring security and centralized control. Other users can only access resources through client computers, as illustrated in Figure 3, which depicts the typical arrangement of computers in a client/server network.

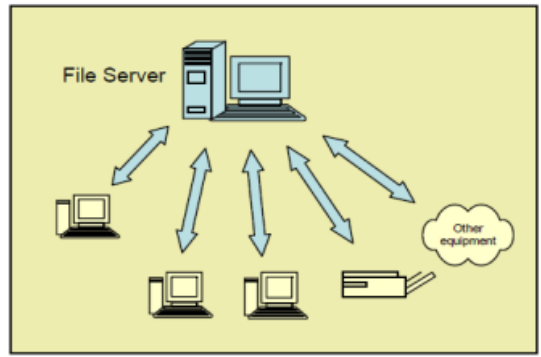

Fig 3: Client-Server Networking

## 2. Types of Networks
### i) Local Area Network

A LAN is generally confined to a specific location, such as floor, building or some other small area. By being confined it is possible in most cases to use only one transmission medium (cabling). This technology is less expensive to implement than WAN because you are keeping all of your expenses to a small area, and generally you can obtain higher speed. They are widely used to connect personal computers and workstations in offices and factories to share the resources. Traditional LANs runs at a speed of 10 to 100 mbps have low delay and make very few errors. Never LANs may operate at higher speed up to 100 mbps.

#### (a) Common Physical Topologies

##### (i) Bus Topology

A bus physical topology is a network arrangement in which all devices connect to a common shared cable, known as the backbone. In this topology, computers (workstations and servers) are attached directly to the backbone using connectors. The backbone is typically a long cable that runs through the network, and it is terminated at both ends to prevent signals from reflecting back after they have passed all devices.

Bus topology was one of the first used topologies to connect computers in a network and is considered one of the oldest forms of network topologies. It is a relatively simple and cost-effective design but can be a failure-prone model, as the failure of the backbone cable can disrupt communication for all devices connected to it.

Most bus topologies allow electric or electromagnetic signals to travel in both directions, allowing for bidirectional communication. A LAN with a bus topology is illustrated in Figure 4, showing how computers are connected to the backbone cable.

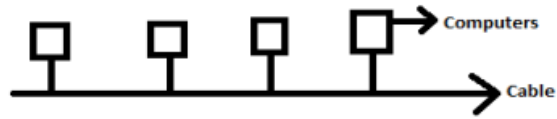

Fig 4: LAN with BUS Topology

**(ii) Ring Topology**

A ring topology is a network configuration where each node is connected to its neighboring nodes, forming a circular pathway for data to travel. Data packets pass around the ring in one direction only. Each device in the ring incorporates a receiver and a transmitter and acts as a repeater, regenerating the signal and passing it to the next device in the ring. This regeneration helps maintain signal integrity, reducing signal degeneration along the ring.

The ring topology emerged as an alternative to the bus topology to overcome its limitations. However, like the bus topology, the ring is also a failure-prone model, as a break in the ring can disrupt communication for all devices in the network.

Ring topologies are well-suited for token passing access methods, where a token is passed around the ring, and only the node that holds the token is allowed to transmit data. This method ensures fair access to the network and helps avoid data collisions.

Despite these advantages, ring topologies are relatively rare in modern networks, with bus and star topologies being more common. However, they are still used in some specialized applications where their unique properties are beneficial.

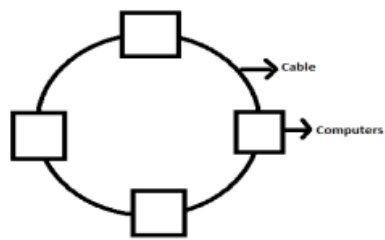

Fig 5: LAN with RING topology

**(iii) Star Topology**

In a star topology, all networked devices are connected to a central device, such as a hub, switch, or multiport repeater, using point-to-point links. This central device acts as a mediator, managing and controlling the flow of data between devices. Unlike bus and ring topologies, where all devices are connected in a linear fashion, star topologies allow for more efficient data transmission and easier troubleshooting.

Each networked device sends electrical or electromagnetic signals up its drop cable to the central device, which then forwards the signal to the appropriate destination device. This arrangement minimizes the chances of signal degradation and collision, making the star topology a reliable and widely used network model.

Star topologies can also be nested within each other to form tree or hierarchical network topologies, providing scalability and flexibility in network design. Overall, the star topology is not only a standard model but also a preferred choice in modern network deployments due to its efficiency and reliability.

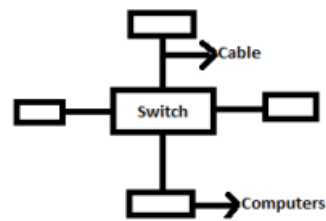

Fig 6: LAN with STAR Topology

**(iv) Mesh Topology**

In a mesh topology, every node is connected to every other node in the network. This creates multiple paths for data to travel, enhancing the reliability and fault tolerance of the network. Mesh topologies can be fully connected, where every node has a direct connection to every other node, or partially connected, where some nodes have direct connections to only a subset of other nodes.

Mesh topologies are highly reliable because if one link or node fails, data can still be transmitted through alternative paths. However, this redundancy comes at the cost of increased complexity and higher implementation and maintenance costs. Mesh topologies are typically used in situations where reliability and fault tolerance are critical, such as in military and critical infrastructure networks.

### (v) Cellular Topology

A cellular topology combines wireless point-to-point and multipoint strategies to divide a geographical area into cells, each representing a specific connection area within the network. Devices within each cell communicate with a central station or switch. These switches are interconnected to route data across the network and provide the complete network infrastructure.

One of the key advantages of cellular topology is its ability to support mobile devices that can roam from one cell to another while maintaining connectivity. This makes cellular topology ideal for wireless communication systems such as mobile networks and Wi-Fi networks in public areas. Each cell can have its own base station or access point, allowing for efficient use of wireless spectrum and minimizing interference between cells.

## ii) Wide Area Network

A Wide Area Network (WAN) is a network that spans a large geographical area, such as a country or continent. It connects multiple Local Area Networks (LANs) that may be separated by significant distances. In a WAN, the network typically consists of numerous cables or telephone lines, each connecting a pair of routers. If two routers that do not share a direct connection wish to communicate, they must do so indirectly.

To achieve communication over long distances, personal computers often use modems to connect to other computers indirectly. WANs play a crucial role in connecting different networks and enabling communication between geographically dispersed locations. A representation of a WAN connecting two different networks is shown in Figure 7.

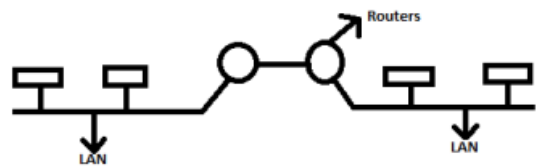

Fig 7:WAN connecting two different networks

### iii) Metropolitan Area Network

A Metropolitan Area Network (MAN) is a network that spans a metropolitan area, typically a city or a large campus. MANs are larger than Local Area Networks (LANs) but smaller than Wide Area Networks (WANs). They are designed to provide connectivity to a specific geographic area and often connect multiple LANs within that area.

MANs are usually owned and operated by a single entity, such as a city government or a large corporation, and are used to interconnect various local networks to enable data and resource sharing. They often use high-speed connections, such as fiber optic cables or wireless technologies, to provide fast and reliable communication within the metropolitan area.

MANs are commonly used to connect office buildings, educational institutions, and government facilities within a city or campus, allowing users to access shared resources and services across different locations.

### iv) Wireless Network

A wireless network is a type of computer network that uses wireless data connections between network nodes instead of physical cables. Devices such as computers, smartphones, and tablets communicate with each other using radio waves. Wireless networks offer flexibility and convenience, allowing users to connect to the network without being constrained by physical cables. However, they can be susceptible to interference and security risks, so proper security measures should be implemented to protect the network and the data transmitted over it.

## 3. Communication Links

Various types and forms of communication medium are:
- Fiber-optic cable.
- Twisted-pair copper wire.
- Coaxial cable.
- Wireless local-area links. (e.g. 802.11, Bluetooth)
- Satellite channel [3].

## 4. Open Systems Interconnection (OSI) Model

The OSI (Open Systems Interconnection) model is a conceptual framework that standardizes the functions of a telecommunication or computing system into seven abstraction layers. Each layer serves a specific purpose and interacts with the adjacent layers to enable communication between devices over a network.

The seven layers of the OSI model are:

- Physical Layer: This is the lowest layer of the OSI model and is responsible for transmitting raw data bits over a physical medium.

- Data Link Layer: This layer is responsible for framing data into frames and ensuring error-free transmission over the physical layer.

- Network Layer: The network layer is responsible for routing and forwarding data packets between different networks.

- Transport Layer: This layer provides reliable, end-to-end communication between devices and handles issues such as flow control and error correction.

- Session Layer: The session layer establishes, maintains, and terminates communication sessions between devices.

- Presentation Layer: This layer is responsible for translating data into a format that is readable by the application layer.

- Application Layer: The application layer provides network services to user applications and is the closest layer to the end user.

The OSI model provides a standardized framework for designing, implementing, and troubleshooting network communication, allowing different systems to communicate effectively regardless of their underlying technologies.

Table 1: Layers of OSI model and their purpose

| Layer | Purpose |
| --- | --- |
| Physical | Network Interface Card, wire and so on. |
| Data Link | Error checking, manages link control, communication with cards. |
| Network | Addressing, traffic, switching. |
| Transport | Handles network transmission |
| Session | Establishes rules for communication, determines synchronization. |
| Presentation | Translator between application and others, redirector, encryption, compression. |
| Application | Interface to network services. |

# CONCLUSIONS

In This research paper has delved into various aspects of computer networks, covering different types of network configurations, networks, communication links, and the Open Systems Interconnection (OSI) model.

We discussed two main types of network configurations: peer-to-peer networks and client/server networks. Peer-to-peer networks are suitable for smaller networks where strict security is not necessary, while client/server networks are more appropriate for larger networks where centralized control and management are important.

In terms of network types, we explored several, including Local Area Networks (LANs), Wide Area Networks (WANs), Metropolitan Area Networks (MANs), and Wireless Networks. Each type serves a specific purpose and has its own characteristics and applications.

Communication links play a crucial role in network connectivity, with different types of links, such as wired, wireless, and satellite links, offering various advantages and disadvantages in terms of speed, reliability, and cost.

Finally, we discussed the OSI model, which is a conceptual framework that standardizes the functions of a telecommunication or computing system into seven abstraction layers. This model helps network designers and developers understand and implement networks more effectively.

In conclusion, this research paper has provided a comprehensive overview of computer networks, covering various aspects such as network configurations, types, communication links, and the OSI model. Understanding these concepts is essential for designing, implementing, and managing efficient and reliable computer networks in today's interconnected world.